\documentclass{iau} 
\usepackage{graphicx}

\title[A new method to derive SFH] 
{A new method to derive star formation histories in dwarf galaxies}

\author[Marius \v{C}eponis, Rima Stonkut\.{e} \& Vladas Vansevi\v{c}ius]   
{Marius \v{C}eponis$^1$, Rima Stonkut\.{e}$^{1,2}$, \and Vladas Vansevi\v{c}ius$^{1,2}$}

\affiliation{$^1$Center for Physical Sciences and Technology, \\ Saul\.{e}tekis av. 3, 10257 Vilnius, Lithuania \\ email: {\tt marius.ceponis@ftmc.lt} \\[\affilskip]
$^2$Astronomical Observatory of Vilnius University, \\ M. K. \v{C}iurlionis st. 29,  03100 Vilnius, Lithuania \\ email: {\tt vladas.vansevicius@ff.vu.lt}}

\pubyear{2018}
\volume{344}  
\jname{Dwarf Galaxies: From the Deep Universe to the Present} 
\editors{A.C. Editor, B.D. Editor \& C.E. Editor, eds.}
\begin{document}

\maketitle

\begin{abstract}
We present a new method to derive 2D star formation histories in dwarf irregular galaxies. Based on multicolor stellar photometry data we have found that in the Leo~A galaxy during the last $\sim$400~Myr star formation was propagating according to the inside-out scenario. Star-forming regions have spread strongly asymmetrically from the center and their present day distribution correlates well with the H\,{\scshape i} surface density maps.

\keywords{galaxies: dwarf, galaxies: irregular, galaxies: individual Leo~A}
\end{abstract}

\firstsection 
\section{Introduction}

A method to consistently determine 2D star formation histories of resolved dwarf irregular galaxies within the Local Volume ($\lesssim$5~Mpc), based on the upper part of their color-magnitude diagrams, is required for the analysis of the Hubble Space Telescope and 8-10~m class ground-based telescope imaging data archives, which are widely available. To test our method, we selected the well-studied dwarf irregular galaxy Leo~A, which is one of the most isolated and ``youngest'' in the Local Group (\cite[Cole \etal\ 2007]{Cole_etal07}). 

\section{Data}

We used the catalog of Leo~A stars (\cite[Stonkut\.{e} \etal\ 2014]{Stonkute_etal14}) measured on $B$, $V$, and $I$ passband CCD images obtained with the Subaru Telescope equipped with the Suprime-Cam mosaic camera. The catalog was cleaned from obvious non-stellar objects and only stars residing within the ellipse of a semi-major axis, $a=3.5'$, and ellipticity, $b/a=0.6$, (\cite[Vansevi\v{c}ius \etal\ 2004]{Vansevicius_etal04}) were selected. We rejected stars fainter than $V=24.0$ because of low completeness. Additionally, 42 objects with $V-I>1.7$, $B-V>1.7$, $B-I>3.3$ were rejected, based on two-color diagram analysis, as foreground stars. Finally, 2539 stars were selected to study the star formation history (SFH) (Fig.\,\ref{fig1}, upper row). Also, a subset of young stars was selected by cutting out the region of the color-magnitude diagram (CMD) not covered by the isochrones younger than 1~Gyr (Fig.\,\ref{fig1}, bottom row). Therefore, 862 stars in total were selected to represent a young population.

For the generation of artificial stellar populations the PARSEC isochrones (release v1.2S + COLIBRI PR16, \cite[Marigo \etal\ 2017]{Marigo_etal17}) were employed. The initial mass function (IMF) was assumed to follow \cite[Kroupa (2001)]{Kroupa01}. The values of foreground interstellar extinction ($E(B-V)=0.021$) and the distance (800~kpc, $m-M=24.5$; $1'\approx230$~pc) to Leo~A were taken from \cite[Cole \etal\ (2007)]{Cole_etal07}.

\section{Method}

\begin{figure}[b]
\begin{center}
 \includegraphics[width=5.1in]{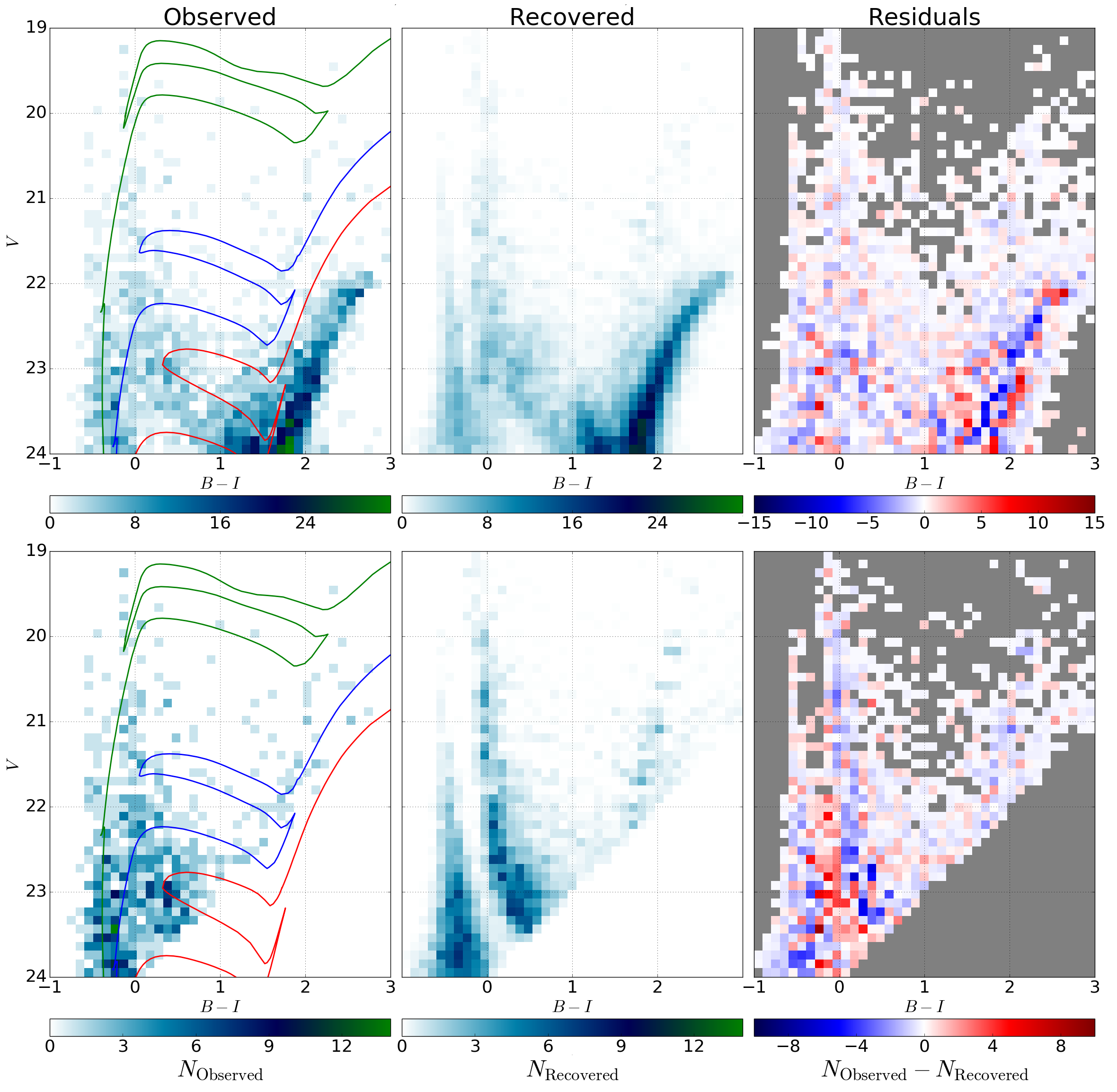} 
 \caption{Hess diagrams of the Leo~A galaxy used for SFH determination (left), synthetic star populations generated according to the recovered SFH (middle), and residuals (right). An upper raw represents all and a bottom raw -- young star populations. In the ``observed" diagrams three isochrones of 50 (green), 200 (blue), 500 (red)~Myr and metallicity of $Z=0.0007$ are plotted. The middle diagram is an average of 10 synthetic populations to smooth out stochastic effects. A number of stars in each bin is color-coded according to the bars at the bottom of each diagram. In the diagrams of residuals (right), bins that have neither observed nor synthetic stars are shown in grey.}
 \label{fig1}
\end{center}
\end{figure}

\begin{figure}[b]
\begin{center}
 \includegraphics[width=5.3in]{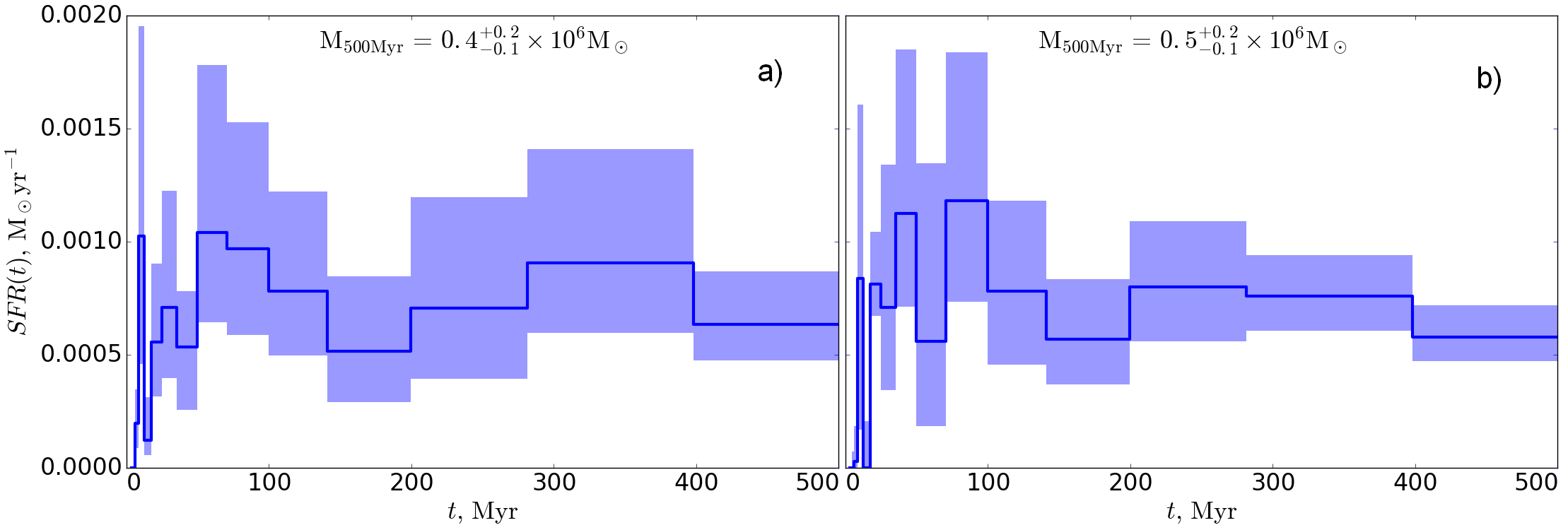} 
 \caption{The derived recent SFHs of the Leo~A galaxy: a) from all stars, $V<24$; b) from young stars, $V<24$. Dark blue lines show the star formation rate at a given time, \textit{SFR(t)}, in units of solar masses per year, $\mathrm{M}_{\odot}\mathrm{yr}^{-1}$. Pale blue areas indicate the scatter limits of the derived \textit{SFR(t)}, corresponding to the lower limits of real SFH accuracy.}
 \label{fig2}
\end{center}
\end{figure}

\begin{figure}[b]
\begin{center}
 \includegraphics[width=5.3in]{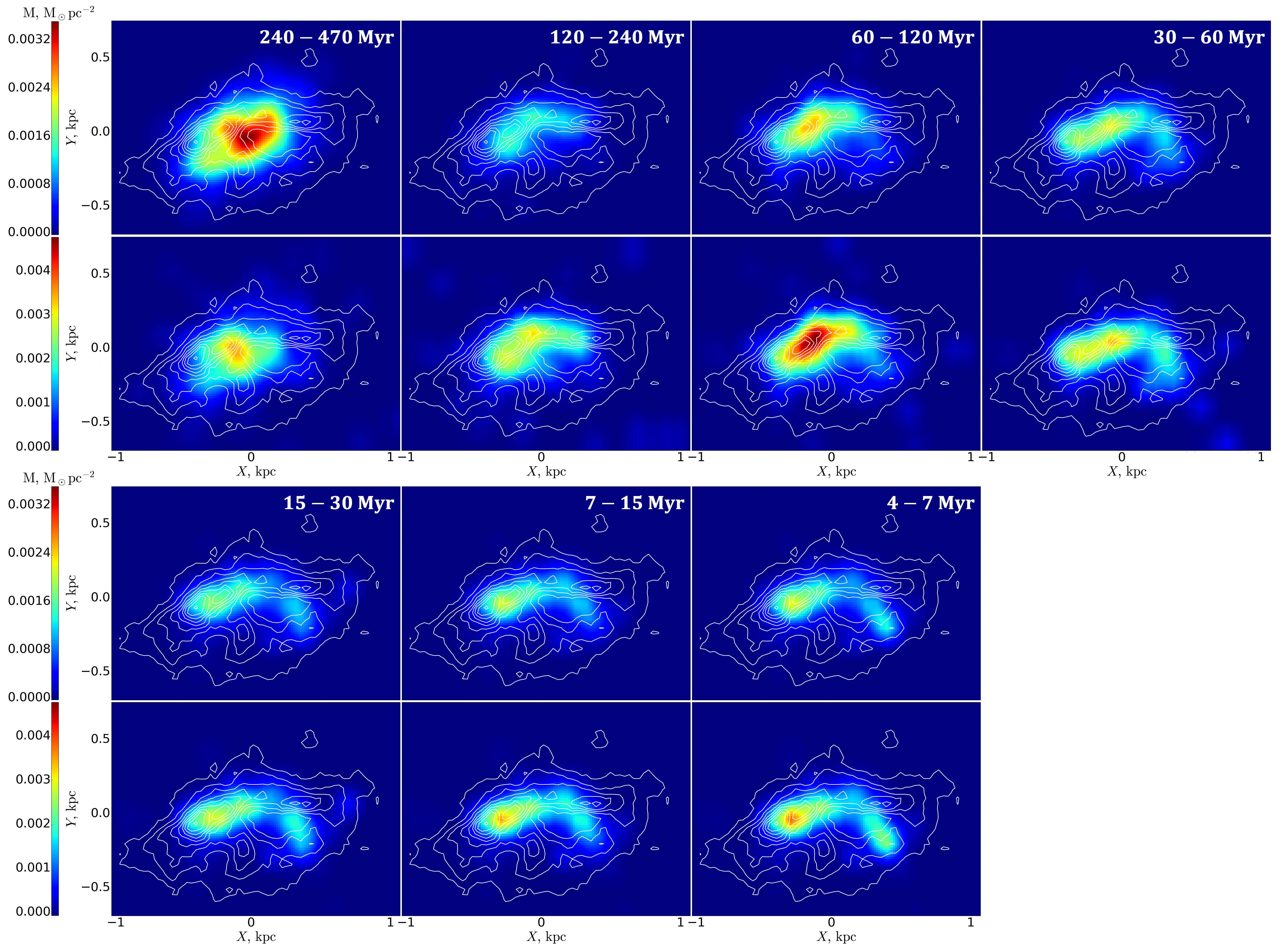} 
\caption{The 2D distributions of stellar mass formed in the Leo~A galaxy derived from the following data sets: (1st and 3rd rows) all observed stars ($V<24$, $N=2539$), upper row in Fig.\,\ref{fig1}; (2nd and 4th rows) selected young stars ($V<24$, $N=862$), bottom row in Fig.\,\ref{fig1}. White contours indicate the present day H\,{\scshape i} column density distribution (\cite[Hunter \etal\ 2012]{Hunter_etal12}). North is up, East is left in all panels.}
 \label{fig3}
\end{center}
\end{figure}

The SFH of Leo~A was determined using a modified two-step fast iterative method described in \cite{Dolphin02}. By comparing binned CMDs (Hess diagrams) of observed stellar populations with artificially generated ones, parameters of synthetic populations are iteratively adjusted to match observations. All synthetic populations are generated taking into account measurement effects determined from the Artificial Star Test (AST) results (\cite[Stonkut\.{e} \& Vansevi\v{c}ius 2015]{Stonkute_Vansevicius15}). As an initial guess, we use a constant star formation rate, $SFR(t,Z)=0.001\,\mathrm{M}_{\odot}/\rm{yr}$, from 4~Myr to 13~Gyr and metallicities $Z\in[0.0001;0.0015]$. At each iteration the 3D Hess diagrams ($B$, $V$, $I$) of synthetic and observed stellar populations are compared and parameters (mass (M), $Z$) of stellar populations of each age are updated:

\begin{equation}
{\rm M}_{i+1}(t,Z)={\rm M}_{i}(t,Z) \frac{\sum_j n_j(t,Z)\,s_j/o_j}{\sum_j n_j(t,Z)},
\label{eq:eq1}
\end{equation}

\noindent here, $s_j$ denotes a number of synthetic stars in a bin $j$, $o_j$ -- a number of observed stars, and $n_j(t,Z)$ -- a number of synthetic stars of a particular age, $t$, and metallicity, $Z$. For the bins which contain only observed stars, $o_j>0$, and no synthetic ones, $s_j=0$, we assume a ratio $o_j/s_j=0.5$. The iterative procedure is repeated until the sum of differences in star numbers $\sum_j o_j-s_j$ converges to some constant value. SFH of Leo~A usually converges after $\sim$100 iterations (Fig.\,\ref{fig1}). To estimate the accuracy of the determined SFH, results from the last 200 iterations are used (Fig.\,\ref{fig2}).

Also, we derive 2D distributions of stellar mass formed during various time intervals (Fig.\,\ref{fig3}). Firstly, for each observed star, a probability to be of particular age, metallicity, and mass is evaluated by comparing photometric data with theoretical isochrones. Photometric errors and a completeness function are estimated from the AST results. Probabilities are calculated for all selected age, metallicity, and stellar mass values. Then, by combining the derived probability distributions of each star, a probability-weighed stellar mass for particular age intervals and sky coordinates is calculated.

\section{Results and Discussion}
We have developed a new method optimized to derive recent ($\lesssim$400~Myr) 1D and 2D SFHs of the resolved dwarf irregular galaxies (rigid body rotation) within the Local Volume ($\lesssim$5~Mpc, significant part of CMD is observed), based on multicolor stellar photometry data. We applied it to derive SFH of Leo~A (Fig.\,\ref{fig2}). A total stellar mass of Leo~A inside the Holmberg radius was estimated at ($3.8\pm1.6)\times10^6\,\mathrm{M}_{\odot}$. SFHs based on the complete and reduced datasets (Fig.\,\ref{fig1}) agree reasonably well, considering they are derived from data sets differing by three times in star number. Three periods of star formation strengthening are observed at ages of $\sim$30, 80, 300~Myr. 

Also, 2D distributions of formed stellar mass (Fig.\,\ref{fig3}) were derived using two data sets: all observed stars (1st and 3rd rows) and only young stars (2nd and 4th rows). Our results show that during the last $\sim$400~Myr star formation in this galaxy followed an inside-out scenario. At the beginning ($\sim$400~Myr ago), a clear star-forming region is visible residing at the center of Leo~A. Over the next 200~Myr it evolved into an asymmetrical shape, which correlates well with the current H\,{\scshape i} surface density distribution. It is worth mentioning that a large hole in the H\,{\scshape i} distribution (to the West from the center of galaxy, Fig.\,\ref{fig3}) has been almost free of star formation for the last $\sim$240~Myr. This suggests that the H\,{\scshape i} hole was formed by multiple generations of stars. However, conclusions should be made with caution since stars could migrate away from their birthplace significantly. Nonetheless, a strong evolution of star-forming regions can be seen clearly.
\\

$Acknowledgements$. This research was funded by a grant No. LAT-09/2016 from the Research Council of Lithuania.

\end{document}